\documentclass[]{spie}  

\usepackage{amsmath,amsfonts,amssymb}
\usepackage{graphicx}
\usepackage[colorlinks=true, allcolors=blue]{hyperref}
\usepackage{orcidlink}
\newcommand{\orcid}[1]{\,\orcidlink{#1}}
\usepackage{booktabs}  
\usepackage{listings}
\usepackage{xcolor}

\usepackage{aas_macros}

\lstdefinelanguage{json}{
    basicstyle=\ttfamily\footnotesize,
    numbers=left,
    numberstyle=\tiny\color{gray},
    stepnumber=1,
    numbersep=8pt,
    xleftmargin=2.5em,
    showstringspaces=false,
    breaklines=true,
    breakatwhitespace=true,
    frame=single,
    rulecolor=\color{gray!50},
    backgroundcolor=\color{gray!5},
    literate=
     *{:}{{{\color{black}{:}}}}{1}
      {,}{{{\color{black}{,}}}}{1}
      {\{}{{{\color{blue}{\{}}}}{1}
      {\}}{{{\color{blue}{\}}}}}{1}
      {[}{{{\color{blue}{[}}}}{1}
      {]}{{{\color{blue}{]}}}}{1},
    string=[s]{"}{"},
    stringstyle=\color{red!70!black},
    comment=[l]{//},
    commentstyle=\color{gray}\itshape,
}

\title{Test Management and Coordination During the Vera C. Rubin Observatory Commissioning and Early Operations Using Zephyr Scale}

\author[a]{Bruno Quint\orcid{0000-0002-1557-3560}}
\author[b]{Tiago Ribeiro\orcid{0000-0002-0138-1365}}
\author[b]{Erik Dennihy\orcid{0000-0003-2852-268X}}
\author[b]{Brian Stalder\orcid{0000-0003-0973-4900}}
\author[a]{David Sanmartim\orcid{0000-0002-9238-9521}}
\author[b, c]{Keith Bechtol\orcid{0000-0001-8156-0429}}

\affil[a]{NSF–DOE Vera C. Rubin Observatory/NSF NOIRLab, Casilla 603, La Serena, Chile}
\affil[b]{NSF–DOE Vera C. Rubin Observatory/NSF NOIRLab, 950 N. Cherry Ave., Tucson, AZ 85719, USA}
\affil[c]{Department of Physics, University of Wisconsin-Madison, Madison, WI 53706, USA}

\authorinfo{Further author information: (Send correspondence to B.Q.)\\B.C: E-mail: bruno.quint@noirlab.edu, Telephone: +56 51 2205 200}

\begin{document} 
\maketitle

\begin{abstract}
The commissioning of the NSF–DOE Vera C. Rubin Observatory required coordinating the planning, design, and execution of hundreds of integration and on-sky tests involving different subsystems and geographically distributed teams. To support this task, we adopted a Jira-native test management tool, Zephyr Scale. The initial use of Zephyr Scale focused solely on system verification and validation. Its use was rescoped to coordinate higher-level tests, and it is still in use in early operations.  

Zephyr Scale allows the creation of Test Cases, which represent individual tests. Each Test Case contains the information needed to execute a test at the summit. This includes a step-by-step script. Every day, Test Cases are grouped into a Test Cycle, which represents the test plan for all tests to be executed that day and that same night. 

We describe the defined workflow for test creation, review, and deployment, which bridges the gap between ideation and on-sky execution within a Test Cycle. We also outline how we write more complex tests as partially automated JSON files consumed by the Scheduler—the system's real-time, constraint-aware observation optimization engine. This integration enables the Scheduler to ingest high-level observing scripts that communicate with subsystems via an abstraction layer to execute common observatory operations, such as slewing, tracking, and data acquisition.

Finally, we summarize the benefits and limitations of using Zephyr Scale, designed initially to coordinate software testing, for large-scale observatory commissioning and early operations.

\end{abstract}



\section{INTRODUCTION}
\label{sec:intro}

The NSF–DOE Vera C. Rubin Observatory formally completed its construction phase, together with the last part of Systems Integration, Testing and Commissioning late in 2025. The observatory operates two telescopes at Cerro Pachón, in Chile: the 8.4\,m Simonyi Survey Telescope, which carries the LSST Camera (LSSTCam)~\cite{LSSTCam}, and the 1.2\,m Auxiliary Telescope (AuxTel), used for atmospheric characterization. Before LSSTCam was installed, the Simonyi Survey Telescope was commissioned on sky with the LSST Commissioning Camera (LSSTComCam)~\cite{LSSTComCam}. LSSTComCam should not be confused with LSSTCam: the two names differ only by the abbreviation for ``commissioning'', so we spell them out in full throughout this paper to keep them distinct. This was a fast-paced phase of the observatory's history with daily changes in procedures and guidelines. During peak commissioning activities, there were more than fifty individuals with different roles and expertise involved in designing, running, or analyzing the results of hundreds of tests. 

Now the observatory is in the early operations stage and we expect to replace the constant changes with a steady routine. However, even in this new era, coordinating tasks at this scale still requires the information to propagate to more than thirty people in different teams, and quickly. Some of the teams involved in the daily coordination are the Observing Specialists, the Rubin Summit Scientists, and the In-Kind Contributors. We refer to these groups collectively as the Observing Team.

The Observing Specialists are the main drivers of the observatory and execute the individual tasks that are part of each night's plan. Rubin Summit Scientists provide technical and scientific support to implement tests and plan the nightly operations. In-Kind Contributors bring expertise on higher-level performance and requirements. Together, Rubin Summit Scientists and In-Kind Contributors design new tests and perform analyses on image quality, stray light, and thermal and environmental performance.

To support coordination at this scale, we adopted Zephyr Scale\footnote{\url{https://smartbear.com/test-management/zephyr-scale/}}, a Jira-native tool originally designed for software testing. During early stages of commissioning, we re-scoped this tool for verification and validation tests. It proved to be useful for organizing and tracking tests, and for executing them step-by-step. Later, it became the primary tool for tests, procedures, and nightly plans. Zephyr Scale was critical for communication and coordination during commissioning and we still use it during the early stages of operations.

In Section \ref{sec:operational_framework}, we describe what tools, including Zephyr Scale, featured in the daily life during commissioning and early operations and how they worked together. In Section \ref{sec:workflows}, we describe the process used to convert any idea into a formal procedure, how the nightly plans converged, and the workflow on a weekly basis. Finally, we close out this work with a summary of the benefits and limitations of Zephyr Scale when used to coordinate tests during commissioning and early operations.

\section{Operational Framework}
\label{sec:operational_framework}

Coordinating and running activities at the summit involves two main aspects: the technical aspect, namely how the Observing Team interacts with the observatory; and the operational reality, namely that no two days are quite alike. We address the technical aspect first. On the technical side, we need to be able to issue low-level commands to control individual components, and run high-level scripts for more complex operations such as sending the telescope to track a target on sky and take images. Finally, we need to be able to tell the observatory to start a survey and it should run by itself. 

The operational reality is that most days differ from each other. Ideally speaking, we want to have a steady routine. But this is not what happens. We have technical issues. Optimization tests. Dedicated investigations. And we need to decide what happens when. On top of that, we need to ensure that any relevant information about each of these activities is properly passed to the Observing Specialists, who are the final consumers of procedures and documentation. They need to have the tools to record the actual results of tests so we can find these results later and perform analysis. 

This section focuses on the technical side: the Observatory Control System and the two mechanisms we use to drive more complex tests, namely JSON BLOCKs and Scheduler Configurations. The coordination side --- how we plan, document, and track tests with Zephyr Scale, and how it works together with other tools such as Slack, Jira, and Confluence --- is covered in Section \ref{sec:workflows}.

\subsection{Vera C. Rubin Observatory Control System}
\label{ssec:obs_control_system}

The Observatory Control System is a Python-based service that relies on a Service Abstraction Layer (SAL) \cite{Mills2016_SAL}, which uses Kafka as the main communication protocol \cite{Ribeiro2024_Kafka}. Each relevant hardware component in the observatory is represented as a Commandable SAL Component (CSC) \cite{Lotz2016_CSC} which works as a link between its low-level controller and the rest of the system.

Users interact with CSCs through two main interfaces. The simplest are SAL Commands, which target a single CSC directly. For more complex tasks, we rely on SAL Scripts — special Python scripts that coordinate multiple CSCs, include status checks, and handle error conditions. Both SAL Commands and SAL Scripts are dispatched through the \texttt{ScriptQueue}, a dedicated CSC, using the LSST Observatory Visual Environment (LOVE)\footnote{\url{https://github.com/lsst-ts/LOVE-manager}} as the front-end interface.

On top of SAL Scripts, we have the Scheduler \cite{Delgado2016_Scheduler}, a dedicated CSC that drives nighttime observations automatically. Its default algorithm is the Feature Based Scheduler (FBS) \cite{Naghib2019, Delgado2016_Scheduler}, a real-time optimization engine that reads the current weather and system conditions, the telescope state, and the history of prior observations to decide where the telescope should be pointed next. The primary goal of the Scheduler is to optimize on-sky coverage to accomplish the Legacy Survey of Space and Time (LSST) \cite{Ivezic2019_LSST}.

The Scheduler can be used with different configurations. The default configuration tells the Scheduler to perform the LSST. However, some very specific tests require live target production following some special criteria. For example, the Active Optics System stability tests, where we point the telescope at a position on sky, track while taking a set of exposures, and go back to this original position. We do this for a long time to isolate the optical aberrations from thermal or atmospheric effects in the image quality. 

Whenever we run any test or task at the Rubin Observatory, we need to send a specific sequence of SAL Scripts to the control system. The primary tool to plan, document, and track all these tests is Zephyr Scale, which is described in more detail in Section \ref{ssec:zephyr_scale}. Some tests are complex enough to require a predefined sequence of SAL Scripts stored in a dedicated file in JSON format called JSON BLOCKs (Section \ref{ssec:json_blocks}), while others require the Scheduler to dynamically select targets following specific criteria, handled through Scheduler Configurations (Section \ref{ssec:scheduler_configurations}).

\subsection{JSON BLOCKs}
\label{ssec:json_blocks}

In some cases, tests require running a large number of scripts or specific sequences of scripts. In such situations, instead of adding each script to the \texttt{ScriptQueue} manually, we store them in a JSON file called a JSON BLOCK. Each JSON BLOCK has three mandatory fields: \texttt{name}, a human-readable identifier; \texttt{program}, a unique identifier used internally by the Scheduler — typically matching the associated BLOCK Jira ticket; and \texttt{scripts}, the ordered list of SAL Scripts and their configurations. Some JSON BLOCKs also expose a limited set of parameters that can be passed at execution time, allowing for minor variations without requiring a new file. Listing~\ref{lst:json_block} shows the basic structure of a JSON BLOCK. 

\begin{lstlisting}[language=json, caption={Basic structure of a JSON BLOCK file.}, label={lst:json_block}]
{
    "name": "BLOCK-T123",
    "program": "BLOCK-T123",
    "scripts": [
        {
            "name": "maintel/slew_and_take_image.py",
            "standard": true,
            "parameters": { ... }
        }
    ]
}
\end{lstlisting}

All JSON BLOCKs are stored in the public \texttt{ts\_config\_scheduler} GitHub repository\footnote{\url{https://github.com/lsst-ts/ts_config_scheduler}} and are loaded into the Scheduler when it is enabled, at which point they are validated as proper JSON and their script configurations are checked against the corresponding schemas. If a JSON BLOCK fails validation, the Scheduler transitions to the FAULT state, and the file must be corrected in the repository and reloaded. The primary interface to execute a JSON BLOCK is the \texttt{add\_block.py} SAL Script, which instructs the Scheduler to read the file and populate the \texttt{ScriptQueue} with the associated scripts and their configurations.

Each JSON BLOCK is directly associated with a Test Case. However, a single Test Case may be backed by more than one JSON BLOCK — for example, when different versions of a test require slightly different sequences. A complete real-world example is provided in Appendix~\ref{app:json_block}. The workflow for creating and deploying a JSON BLOCK — from the initial idea to its availability at the summit — is described in Section~\ref{ssec:each_test}. JSON BLOCKs work well when the sequence of scripts is fixed. When a test requires dynamic target selection instead, the Scheduler is driven by a Scheduler Configuration, described in the next section.

\subsection{Scheduler Configurations}
\label{ssec:scheduler_configurations}

As introduced in Section~\ref{ssec:obs_control_system}, the Scheduler is a dedicated CSC that runs the FBS by default. A detailed description of the FBS is beyond the scope of this paper and is covered extensively elsewhere \cite{Naghib2019}; here we focus on how Scheduler Configurations are used in the context of commissioning tests.

A Scheduler Configuration typically involves a combination of Python, YAML, and JSON files stored in the \texttt{ts\_config\_scheduler} repository. Together, these files define the observing strategy the FBS will follow: which regions of the sky to prioritize, which telemetry values to monitor and their acceptable limits, and the cadence of observations. For each target selected by the FBS, the Scheduler executes an associated JSON BLOCK that defines the exact sequence of SAL Scripts to run at that position.

Scheduler Configurations are used for two broad categories of tests. In some cases, the Scheduler is used purely for convenience, even when the target selection is not dynamic. For example, in Active Optics System stability tests, the telescope is commanded to a fixed azimuth and elevation, aligned, and then a triplet of exposures is taken before returning to the initial position — a cycle that repeats to isolate optical aberrations from thermal or atmospheric drift. In other cases, the Scheduler is genuinely needed for dynamic target selection, such as observations that require special parameter configurations or dedicated sequences of scripts for each target, chosen in real time based on current conditions.

\section{Workflows}
\label{sec:workflows}

At the center of these workflows is Zephyr Scale, the Jira-native tool we use to plan, document, and track every test. We describe it first and then the processes built around it.

\subsection{Zephyr Scale}
\label{ssec:zephyr_scale}

Zephyr Scale is a Jira-native test management software designed to support software teams to plan and run software tests. We re-scoped the application to perform verification tests in the early stages of the System Integration and Commissioning phase of the Rubin Observatory. Zephyr brings the concept of Test Cases, Test Cycles, Test Plans, and Test Executions (Table~\ref{tab:zephyr_entities}). Most of these expressions are originally used in software testing.

\begin{table*}[!ht]
\centering
\caption{Zephyr Scale entities used at the Rubin Observatory. The Test Plan entity is supported by Zephyr Scale but is not adopted in our workflow.}
\label{tab:zephyr_entities}
\begin{tabular}{llll}
\toprule
\textbf{Entity} & \textbf{Represents} & \textbf{Lifetime} & \textbf{Example} \\
\midrule
Test Case      & Reusable test procedure           & Persistent; versioned             & \texttt{BLOCK-\textbf{T}123} \\
Test Cycle     & Dated plan grouping Test Cases    & One operational day               & \texttt{BLOCK-\textbf{R}456} \\
Test Execution & One run of a Test Case in a Test Cycle & Frozen snapshot\textsuperscript{a} & \texttt{BLOCK-\textbf{E}789} \\
Test Plan      & Higher-level grouping of Test Cycles   & Not adopted\textsuperscript{b}    & \texttt{BLOCK-\textbf{P}xxx} \\
\bottomrule
\addlinespace
\multicolumn{4}{l}{\footnotesize\textsuperscript{a}A Test Execution can be refreshed against its parent Test Case only if explicitly requested.} \\
\multicolumn{4}{l}{\footnotesize\textsuperscript{b}We use Confluence pages for weekly plans instead; they offer more flexibility for formatting and discussion.} \\
\end{tabular}
\end{table*}

Test Cases represent individual tests. In the context of the Rubin Observatory's commissioning and early operations, a Test Case can represent standard procedures or dedicated tests. Each Test Case has a unique identifier, or key, and other information linked to it. A Test Case key uses the same prefix associated with a Jira Space where it was enabled. For example, at the Rubin Observatory, we have a Jira Space called BLOCK, named after ``Observing Blocks'' which were, at some point, Jira tickets with instructions for tests on sky. Tickets created in this space have the \texttt{BLOCK-[1-9]\textbackslash{}d*} format (e.g., BLOCK-123). Test Cases have the \texttt{BLOCK-T[1-9]\textbackslash{}d*} format (e.g., BLOCK-T123). 

Each Test Case contains high level information such as a name, a high-level description and a field for pre-conditions. It also contains several fields to hold more specific metadata such as the priority, status, expected time, etc. The most important part of a Test Case for night time operations is the Test Script, which is a collection of steps containing what needs to be done, the expected results and other information like which SAL script do we want to run and what configuration we want to apply. Depending on the complexity of the test, a Test Case may be backed by a JSON BLOCK or a Scheduler Configuration, described in Sections \ref{ssec:json_blocks} and \ref{ssec:scheduler_configurations} respectively.

Similarly to Test Cases, the Test Cycle terminology is used originally in software testing to represent a collection of tests that will be part of a software lifecycle. In the context of Rubin Observatory testing, Test Cycles represent a \textbf{plan to be executed on a particular date} (either during the day or during the night). Here, a Test Cycle key can be represented by the regular expression \texttt{BLOCK-R[1-9]\textbackslash{}d*} (e.g., BLOCK-R123). 

Every day we create a new Test Cycle containing a collection of Test Cases representing the tasks we need to perform at the summit. Every time a Test Case is added to a Test Cycle, Zephyr creates a new Test Execution. One can interpret a Test Execution as a single instance of a Test Case associated with a Test Cycle. Like the others, the key identifier for a Test Execution is unique and is represented by the regular expression \texttt{BLOCK-E[1-9]\textbackslash{}d*} (e.g., BLOCK-E123). 

A Test Execution is a snapshot of a Test Case. Even if a Test Case is updated, the Test Execution will keep all the information from its associated Test Case when it was executed, unless explicitly asked to be updated. This is a useful way to track how a Test Case might evolve in time. A Test Execution also holds the status of individual steps of a Test Case, with the option of adding comments. 

Finally, Zephyr Scale also offers Test Plans, which are meant to group multiple Test Cycles under a higher-level planning structure. However, we did not adopt Test Plans in our workflow. Instead, the team found it more natural to write weekly plans in Confluence pages, which offered more flexibility in formatting and discussion.

\begin{figure}[t]
\centering
\includegraphics[width=\textwidth]{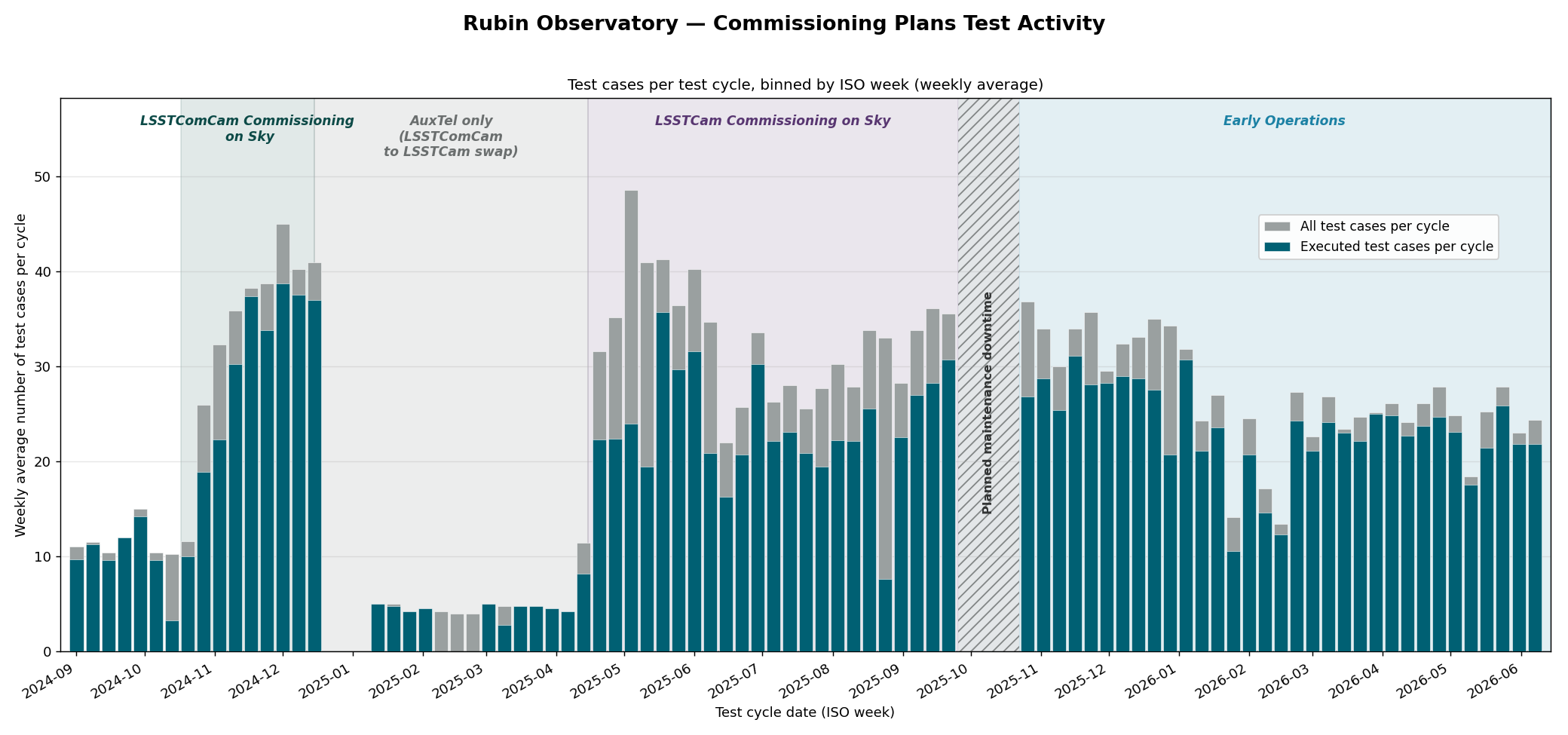}
\caption{Weekly average number of Test Cases per Test Cycle in the BLOCK project's Commissioning Plans, from the start of LSSTComCam Commissioning on Sky through early operations. Each bar is the mean over the Test Cycles created in that ISO week: grey bars count all Test Cases added to a Test Cycle and teal bars count those actually executed (not left in the \textit{Not Executed} state), so the grey portion above each teal bar is the un-executed remainder. Shaded spans mark the main campaigns and the planned maintenance downtime. The data span 526 Test Cycles and 680 distinct Test Cases.}
\label{fig:timeline}
\end{figure}

The scale of adoption gives a sense of how central Zephyr Scale became to daily operations. The statistics reported here, including those shown in Figures~\ref{fig:timeline} and~\ref{fig:steps}, are a snapshot taken in mid-2026 and continue to grow as operations proceed. As of that date, the BLOCK project had accumulated 680 distinct and valid Test Cases, grouped into 526 dated Test Cycles, one per operational day. Adding a Test Case to a Test Cycle creates a Test Execution, and re-running it adds more; in total the project had recorded 15{,}768 Test Executions, of which 12{,}830 (81\%) were actually run. The remaining 2{,}938 were left in the \textit{Not Executed} state --- mostly placeholders carried over when a Test Cycle is cloned from a template or from a previous day rather than tests that were genuinely performed. Figure~\ref{fig:timeline} shows the weekly average number of Test Cases per Test Cycle across the main commissioning and early operations campaigns, with a gap during the planned maintenance downtime. The grey bars count all Test Cases added to a Test Cycle and the teal bars count those actually executed (not left in the \textit{Not Executed} state); the two track each other closely, the grey remainder being tests that we were hoping to run or placeholders that were cloned into a Test Cycle but never run.

Since each Test Cycle is the plan for a single operational day, these counts are also the number of tests carried out on a given night. On a typical night the team executed a few tens of Test Cases: across all 526 nightly plans the median number of Test Cases added to a Test Cycle is 28, of which a median of 23 were executed. This nightly count tracked the observatory's activity. Averaged over Test Cycles in which at least one Test Case was run, it was 32.8 per night (up to 56) during LSSTComCam Commissioning on Sky, dropped to 5.1 while the Simonyi Survey Telescope's camera was swapped from LSSTComCam to LSSTCam and on-sky work continued only on AuxTel, and recovered to 24.3 (up to 46) during LSSTCam Commissioning on Sky and 23.6 (up to 37) in early operations.

Two features of Figure~\ref{fig:timeline} are worth a comment. First, the un-executed remainder is not only material carried over from previous days. The gap is widest during LSSTCam Commissioning on Sky, where it reflects how ambitiously we planned: we routinely added more Test Cases to a Test Cycle than could realistically be executed in a single night. Second, a reader might expect the number of Test Cases per Test Cycle to fall as the observatory settles into operations, yet it stays roughly level. What changed is the mix rather than the volume. Dedicated engineering tests have become less frequent, but the number of subsystems and standard procedures has grown, so fewer engineering tests are offset by more routine daily procedures, each of which is itself captured as a Test Case.

When several individuals are involved, the way we use the tools to plan, coordinate and run activities is as important as the tools themselves. Similar to any large company, most of the plans are discussed in meetings by a small number of individuals. The discussions are followed up in Slack\footnote{\url{https://slack.com}}, together with all the other asynchronous communication throughout the observatory. This includes changes in the plans or in individual processes. 

We present the workflows to create/update individual tests, plan for a night, or set the weekly priorities in the subsections below. For all of them, the starting point will always come from a meeting, go to messages and discussion via Slack, and then be consolidated via Test Cases, Test Cycles or weekly notes in Confluence. 

\subsection{Ideas into Test Cases}
\label{ssec:each_test}

Any functional test, verification test, observation, or standard procedure is now represented as a Test Case. This allows easier communication since we have a unique identifier that represents each Test Case. A Test Case is accessed via web interface and contains five main tabs: Details, Test Script, Traceability, Execution, and History. 

The Details tab holds high-level information such as the Test Case's name, its objective, and preconditions. It also contains other metadata such as the Test Case's priority, status, owner, estimated run time. Zephyr Scale allows adding extra fields to record specific information, such as if a Test Case must run on sky or not. 

The Test Script tab holds the step-by-step procedure to run the associated Test Case. Zephyr Scale allows several ways to represent a Test Script. In our case, we use the ``Step by Step'' version. As the name says, this format of describing a Test Case contains several steps. Each step contains multiple fields such as Step, where we add information about what needs to be done. In most cases, a short description is sufficient. But in other specific tests, we need to extend the text here with links to other pages or dashboards for monitoring. Each step also contains custom fields where we store the name of a SAL script to be executed and the SAL script configuration for that same script. 

The Traceability tab holds links to external pages and to Jira tickets linked to this specific Test Case. Although this is an extremely useful feature, it has not been used in our workflow. The Execution tab lists all the executions of this Test Case. Remember that a Test Execution is automatically created when a Test Case is added to a Test Cycle and, after that, several other executions can be manually created. The Execution tab shows information including the Test Execution status, the end date, and who executed it. 

Finally, the History tab shows every modification in the Test Case with a time stamp, the person who modified it, the field modified, the old and the new value. This allows tracking changes in the Test Case. It is important to note that refreshing a Test Execution to reflect updates in its associated Test Case may cause loss of information if the changes are too large or affect the Test Case's steps. This is an awkward behavior of the system that we learned to work around; it also made us cautious about editing a Test Case while more than one person was working on it at the same time.

Test Cases can have multiple versions, too. This feature brings some benefits. For example, it is easy to change between versions to have Test Executions with different parameters or when most of the steps are similar but with some key differences. However, Zephyr Scale lacks an easy way to compare versions. This becomes a major challenge when a Test Case has too many versions. Most Test Cases only have a few versions --- about 82\% exist in a single version --- but a minority accumulate many: 15 Test Cases (about 2\%) have more than five versions, and one has as many as 19. For these Test Cases with many versions, we need an external tool, such as a page in Confluence, to track all the differences.
Another low level detail regarding versions involves the URL that one can use to access a Test Case. The base URL containing the Test Case's key will only take you to the last version of this Test Case. Once a new version is created, the URL now requires the hash for this specific Test Case and its version, making it hard to integrate with other external tools.

\begin{figure}[t]
\centering
\includegraphics[width=\textwidth]{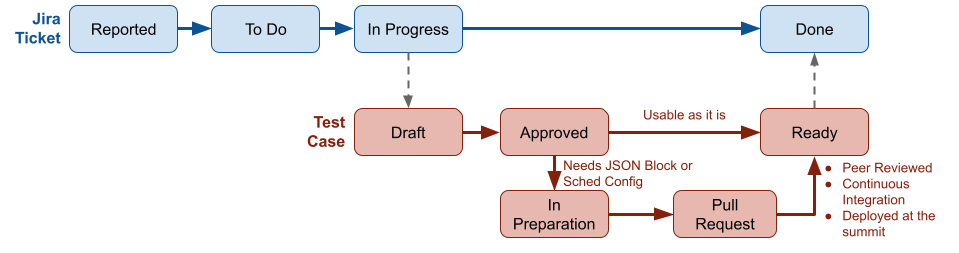}
\caption{Lifecycle of a Test Case backed by a JSON BLOCK or Scheduler Configuration. The Jira ticket and the Test Case progress in parallel, with the ticket closing as Done once the Test Case reaches Ready.}
\label{fig:test_case_workflow}
\end{figure}

Test Cases that involve tens of steps are typically backed by a JSON BLOCK or a Scheduler Configuration rather than spelled out step by step in Zephyr Scale. Any request to create or update either of these usually starts in Slack, in the \texttt{\#rso-observing-blocks} channel, and is formalized via a Jira ticket. Figure~\ref{fig:test_case_workflow} shows how we use the ticket and the Test Case statuses to track the work.

The Jira ticket starts at the ``Reported'' status. Once we identify who will work on the task, we assign them to the ticket and its status becomes ``To Do''. When the assignee actually begins work, they move the ticket to ``In Progress'' and create the Test Case in ``Draft'' status. 

In Draft, the assignee usually starts with a high-level description and slowly improves it as more information becomes available. The Test Case remains in this state until the following criteria are met:

\begin{itemize}
    \item The Details tab contains a high-level description with context and information for people unfamiliar with the test to understand the overall procedure and goals;
    \item The steps in the Test Script are filled up. We strongly encourage keeping tests short, ideally under ten steps, since longer tests are harder to follow during execution. This is usually the case, but there are exceptions.
    \item Complex steps with many commands will be converted into JSON BLOCKs or Scheduler Configurations. The description of the step must contain information enough for people to implement any of these versions. 
    \item Any external resources should be linked to the Test Case in the Traceability tab. This includes Confluence pages, Jira tickets for data analysis, tickets for eventual known issues, tickets containing requirements, and links to associated JSON files.
\end{itemize}

\begin{figure}[t]
\centering
\includegraphics[width=\textwidth]{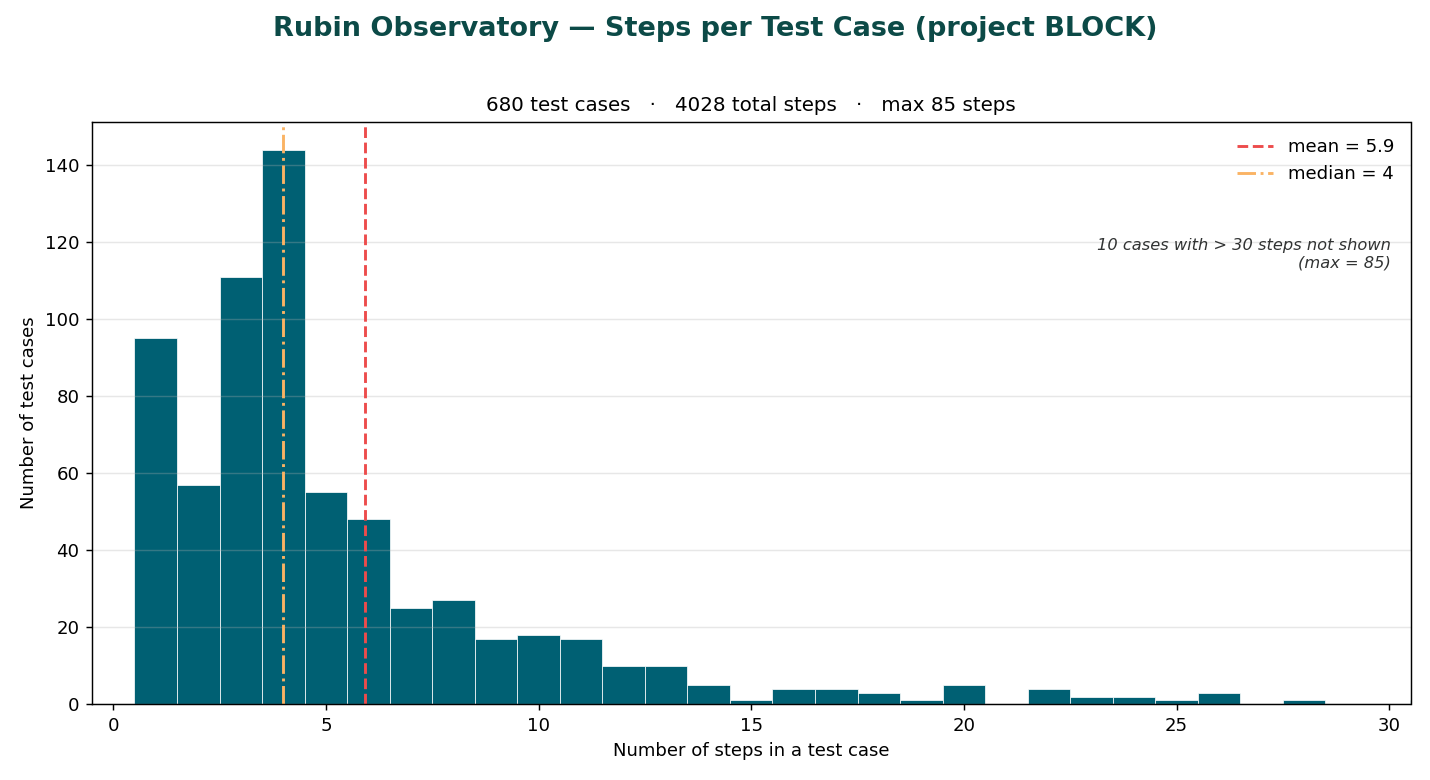}
\caption{Distribution of the number of steps per Test Case across the 680 Test Cases in the BLOCK project. The dashed and dash-dotted lines mark the mean (5.9) and median (4) steps, respectively. The histogram is capped at 30 steps for readability; the 10 Test Cases with more than 30 steps are not shown.}
\label{fig:steps}
\end{figure}

Figure~\ref{fig:steps} shows how well this guideline holds in practice. Across all 680 Test Cases, the Test Scripts have a median of 4 steps and a mean of 5.9. The distribution peaks around four steps and is strongly right-skewed, with most Test Cases at low step counts and a long tail extending toward higher ones: about 88\% of Test Cases have ten steps or fewer, in line with the recommendation to keep tests short, and only 12\% exceed ten steps. A small tail of complex tests runs much longer, with 10 Test Cases exceeding 30 steps. We consider these outliers: a test this long is hard to follow during execution and should instead be broken up or offloaded to a JSON BLOCK or a Scheduler Configuration.

Once the Test Case is mature enough, someone in the Observing Team will review it. Once they claim the Test Case is complete, they set the status as ``Approved''. From here, the original assignee picks up the Test Case again. If the Test Case is ready to use at the summit as it is, they set the Test Case to ``Ready''. Otherwise, if the Test Case requires a JSON BLOCK or a Scheduler Configuration, they set the Test Case as ``In Preparation'' instead.

Both JSON BLOCKs and Scheduler Configurations live in the \texttt{ts\_config\_scheduler} repository. We perform work on any of these files inside the repository in a new branch named \texttt{ticket/\{ticket or Test Case key\}}. Once the changes are deemed complete, the assignee creates a Pull Request to merge the branch into the main branch. The Pull Request now passes through Continuous Integration using Jenkins and a peer review. Once all the checks and reviews pass, the branch is merged and the Pull Request is closed.

Finally, the Summit Scientist deploys the changes to the summit control system and the test is now available to be executed. The assignee updates the Test Case status to ``Ready'' and the Test Case is finally ready to be used in operations. At this point, if there was any Jira Ticket linked to this Test Case, it can be marked as ``Done''.

\subsection{Planning an observing night}
\label{ssec:nightly_plan}

The nightly workflow also starts on Slack. Each afternoon, the day before the night being planned, a Slack Workflow, or a bot, publishes a message to a channel called \texttt{\#rso-test-planning}, typically around 17 h Chilean time. The exact times quoted throughout this workflow are approximate and shift somewhat over the year with the length of the night. This channel is where we discuss short and long term plans, priorities, changes, alternatives, etc. Different stakeholders reply to this automatic message with their requests for tests, following the week's priorities (see Section \ref{ssec:weekly_priorities}).

Early the next morning, before the shift of Observing Specialists starts at around 9 a.m., a Summit Support Scientist playing the role of Test Planner creates a new Test Cycle. This Test Cycle is based on a template with routine activities. Since weekdays and weekends are different, we have one template for each scenario. 

During the day, the Test Planner will collect information across the organization about the observatory's status, requests for tests, and any other existing constraint that might affect nightly operations. They will collect any request in the thread started by the bot above and add the corresponding Test Cases to the Test Cycle, only if the Test Case has the ``Ready'' status.

The number of channels and threads where we discuss plans, changes, and strategies is so large that it is unrealistic to expect the Test Planner to track every place. Instead, our expectation is that any information that might affect night time operations needs to reach the Test Planner using the Slack thread started above. This information is recorded as a narrative set of instructions in the Details tab of the Test Cycle. 

The Test Cycle must be complete by late afternoon, around 16:30, in time for the Evening Tailgate Meeting, where the Test Planner and the Observing Specialists connect to review the night plan together. The Observing Specialists at the summit summarize the current observatory's state and what are the constraints that are likely to be applied on that night. The Test Planner summarizes what is in the Test Cycle and goes through all the deviations for that night.

\subsection{Weekly priorities}
\label{ssec:weekly_priorities}

Once a week, on Wednesdays, we have a meeting to discuss the priorities of the week. This is a short meeting that consolidates the status of the execution of Test Cases, the results, analysis and next steps that are digested in other places. It includes discussions on critical aspects of the observatory image quality and performance. In some of these discussions, we raise the need for new Test Cases which are tackled offline.  

Priorities are chosen by weighing several inputs against the observatory's higher-level goals. We consider the results and analysis from the previous week, the current state and known issues of each subsystem, the readiness of the Test Cases involved, and external constraints such as the season, the expected weather, and the availability of hardware and personnel. A test that unblocks downstream work, addresses a recurring failure, or feeds an ongoing analysis is favored over one that can wait. Because dedicated tests compete with survey time, the meeting also balances how much of each night should be spent on optimization versus routine operations.

Our expectation for the near future is that we only run the observatory in survey mode, supporting LSST. However, we are still in an early operations and optimization phase. So there are still opportunities to improve image quality and night time efficiency. The main focus of this meeting is to evaluate which tests can teach us more about the two issues above so we can improve. These weekly decisions are the cadence through which we steer the longer-term objectives of commissioning and early operations: each week's list is a small step toward higher-level milestones in image quality and nighttime efficiency, and revisiting it every week lets us course-correct as results come in. As the observatory stabilizes and the number of nonstandard Test Cases decreases, we expect this same weekly review to shift from planning dedicated tests toward tracking the statistics of routine operations, which will itself become a source of information about performance.

The outcome of this meeting is a list of Test Cases, ready or not, to be executed between the meeting and the same meeting next week.

\section{Conclusion}
\label{sec:conclusion}

Working with a team the size of the Rubin Observatory's commissioning effort means that communication and coordination are a major challenge, and there is no such thing as an ideal tool for the job. In a fast-paced and chaotic environment like commissioning, there is also no time to build one. Reality forces you to use the tools at hand, even when they were designed for a different purpose. Zephyr Scale is one of those tools.

Because it lives inside Jira, Test Cases plug into the project tracking we already use. Linking a Test Case to a ticket, searching across them, or referencing one from elsewhere in Jira happens naturally. The Test Case / Test Cycle / Test Execution split also maps well onto how we actually work: a Test Case is a reusable procedure, a Test Cycle is a night's plan, and a Test Execution is the record of what was attempted and what happened, frozen in time. Both Test Cases and Test Cycles expose a History tab that records every change, field by field, with author and timestamp. This matters when we go back weeks later to understand why a result looked the way it did.

There are also a few things we wish worked better in Zephyr Scale. There is no built-in way to compare two versions of a Test Case side by side, so we keep Confluence pages on the side to track the changes that matter. This is harmless when a Test Case has a few versions, but when one has as many as nineteen versions, tracking the differences becomes tricky. A lightweight external tool that exposes Test Case versions and their diffs through a stable API would let us build this comparison ourselves, without giving up the Jira-native foundation that has worked so far. Test Cycles also tend to accumulate stale instructions. The temporary guidelines and conditions we write into the Details tab of a Test Cycle are meant for a single night, but they easily outlive their relevance and get carried into later Test Cycles, where they can mislead. The problem of stale information is real, and we address it manually: as part of preparing each Test Cycle, the Test Planner reviews every temporary guideline and condition in the Details tab and removes the ones that no longer apply, so that what remains is only what is relevant for that night.

The URL of a Test Case silently points to its latest version. Once a new version is created, accessing an earlier one requires a version-specific hash in the URL, which is hard to construct programmatically. As a result, a link that was stable yesterday can show different content today, which breaks small integrations and old Confluence references. Finally, all of Zephyr Scale's box text fields — step text, descriptions, preconditions, and the rest — offer limited formatting, and what is there is awkward to use. This leads to inconsistent styles across authors and makes it hard to embed structured information that other tools could read later. Richer formatting such as Markdown would reduce the temptation to bury structured content in free text. None of these are show-stoppers on their own, but they add up across hundreds of Test Cases.

Some of the issues we face are not really about the tool, but about how people use it. \textbf{When a workflow is too complex, people take shortcuts.} That leads to inconsistency. This means that people start losing trust in the process and in the information passed, leading to inefficiencies, time wasted in discussions, and even lost observing time at night. This is also why the information in Test Cases and Test Cycles has to be reliable. For example, when a Test Case has an Approved or Ready status, that has to mean exactly what the workflow says it means. People downstream, including the Test Planner building that night's Test Cycle, act on it without re-reading the Test Case in detail.

Looking ahead, the improvement we would most like to see is a tighter connection between Zephyr Scale and the rest of the observatory's data infrastructure. A Test Execution allows recording what was done from the observer's perspective: which steps were carried out, what was observed, and any comments along the way. The matching technical record lives in the \texttt{ScriptQueue} logs, the Engineering and Facilities Database (EFD), the Consolidated Database (ConsDB), and the Observatory Logging Environment (OLE). The link between the two is currently manual: someone has to know the time window of a Test Execution, find the corresponding visits or telemetry, and bring them together for analysis. If creating a Test Execution automatically captured the \texttt{ScriptQueue} runs it triggered, along with the visits and telemetry they produced, the execution record would become a real entry point into our operational data. Going one step further, LOVE could surface the active Test Execution alongside the controls that drive it, so that the operator sees the procedure and the system response in the same place. None of this requires Zephyr Scale itself to change. It only requires a thin layer between Zephyr Scale, LOVE, and the SAL-based control system that fills in the metadata we currently capture by hand.

Zephyr Scale was meant to be temporary. When commissioning started, the expectation was that we would move to pure Scheduler-based operations and retire the test management layer. That has not happened, and we do not expect it to in the next year or two. Some of this is inertia, but there is also a real reason: Zephyr Scale gives us something that plain documentation does not.

The ability to mark individual steps as executed, along with comments and results, creates an immutable record that is invaluable for operations. A checklist in a Confluence page is just text; a Test Execution is a record that connects procedure to outcome. It is tempting to think a Confluence page with checkboxes could do the same, but it does not scale. On an average night we run a few tens of Test Cases --- a median of 23 executed Test Cases across the Test Cycles we examined --- and a Test Case can have up to ten steps, so a single night can involve a few hundred checkboxes to tick off. The next night we would have to reset all of them, and it is not even obvious how to start a fresh night: clone the page, copy and paste it, or build it from scratch. Zephyr Scale solves this for us by creating a new Test Execution each time a Test Case is added to a Test Cycle, giving every night its own clean, time-stamped record with no manual bookkeeping. We did not find another tool that does something similar.

The ideal tool to replace it will probably never arrive. Integration with LOVE and the \texttt{ScriptQueue} remains on the wishlist, but that is more a matter of resources and prioritization than a limitation of Zephyr Scale itself.

What actually makes the tool work is not the tool itself, but the discipline we put around it. A procedure has to be short enough to be followed and strict enough to be trusted. Zephyr Scale helps us with the first part. The discipline around statuses, reviews, and how we manage Test Cycles is what gets us to the second.

\acknowledgments

This material is based upon work supported in part by the National Science Foundation through Cooperative Agreements AST-1258333 and AST-2241526 and Cooperative Support Agreements AST-1202910 and 2211468 managed by the Association of Universities for Research in Astronomy (AURA), and the Department of Energy under Contract No.~DE-AC02-76SF00515 with the SLAC National Accelerator Laboratory managed by Stanford University. Additional Rubin Observatory funding comes from private donations, grants to universities, and in-kind support from LSST-DA Institutional Members.

The authors thank the Rubin Observatory Observing Specialists, the Summit Support Scientists, and the In-Kind Contributors whose daily use of the workflow described in this paper shaped both the tool and the procedures around it. We also thank the Telescope
and Site, Camera, and Data Management teams for their continued collaboration during integration, commissioning, and early operations.

The authors gratefully acknowledge the use of Atlassian Jira, Atlassian Confluence, SmartBear Zephyr Scale, and Slack as the software infrastructure underpinning the workflows described here.

Anthropic's Claude was used to assist with editorial review, reference discovery, and structural feedback during the preparation of this manuscript; the authors are responsible for all content and conclusions.

\appendix
%
%

\section{Example Observing Block: BLOCK-T70}
\label{app:json_block}

This appendix presents a representative observing block (BLOCK) used at the Vera C. Rubin Observatory to coordinate multi-step operations through the \texttt{ScriptQueue}. Observing blocks are JSON documents that describe a sequence of SAL Scripts to be executed, together with their parameters and an optional configuration schema for runtime customization. They are loaded into the Scheduler CSC and dispatched to the appropriate \texttt{ScriptQueue} for
execution.

The example shown in Listing~\ref{lst:block_t70} is BLOCK-T70, a simple block that exercises the Main Telescope Dome (\texttt{MTDome}) shutter. The block issues an \texttt{openShutter} command, waits for a configurable duration, stops the shutter motion on a specified subsystem, waits briefly, and finally issues a \texttt{closeShutter} command. The full sequence is composed of five SAL Script invocations: three \texttt{run\_command.py} calls and two
\texttt{sleep.py} calls.

The structure of the JSON document follows the conventions adopted across the \texttt{ts\_config\_scheduler} repository\footnote{\url{https://github.com/lsst-ts/ts_config_scheduler}}:

\begin{itemize}
    \item \texttt{name}, \texttt{id}, and \texttt{program} identify the block and link it to its Jira ticket counterpart.
    \item \texttt{constraints} specifies any execution constraints (empty in this case, since the block is intended for engineering use rather than survey scheduling).
    \item \texttt{scripts} is the ordered list of SAL Scripts to execute, each with its own \texttt{name}, a \texttt{standard} flag indicating whether the script is part of the standard or external script repository, and a \texttt{parameters} object passed to the script at runtime.
    \item \texttt{configuration\_schema} embeds a JSON Schema (in YAML form) describing the configurable parameters of the block. In BLOCK-T70, the only configurable parameter is \texttt{sleep\_for}, which controls the delay between opening and closing the shutter, defaulting to 10 seconds. Parameters are referenced inside \texttt{scripts} using the \texttt{\$parameter\_name} substitution syntax.
\end{itemize}

\begin{lstlisting}[
    language=json,
    caption={Observing block BLOCK-T70, used to exercise the Main Telescope
    Dome shutter. Source:
    \href{https://github.com/lsst-ts/ts_config_scheduler/blob/eadd7bdaf809e87aa05cf0d8f9ea087e515d42e2/Scheduler/observing_blocks_maintel/BLOCK-T70.json}{ts\_config\_scheduler/Scheduler/observing\_blocks\_maintel/BLOCK-T70.json}.},
    label={lst:block_t70}
]
{
  "name": "BLOCK-T70",
  "id": "4b0b35ff-5471-42be-8c22-eaf9b8d652af",
  "program": "BLOCK-T70",
  "constraints": [],
  "scripts": [
    {
      "name": "run_command.py",
      "standard": true,
      "parameters": {
        "component": "MTDome",
        "cmd": "openShutter"
      }
    },
    {
      "name": "sleep.py",
      "standard": true,
      "parameters": {
        "sleep_for": "$sleep_for"
      }
    },
    {
      "name": "run_command.py",
      "standard": true,
      "parameters": {
        "component": "MTDome",
        "cmd": "stop",
        "parameters": {
          "subSystemIds": 4
        }
      }
    },
    {
      "name": "sleep.py",
      "standard": true,
      "parameters": {
        "sleep_for": 1
      }
    },
    {
      "name": "run_command.py",
      "standard": true,
      "parameters": {
        "component": "MTDome",
        "cmd": "closeShutter"
      }
    }
  ],
  "configuration_schema": "$schema: http://json-schema.org/draft-07/schema#\ntitle: BLOCK-T70 configuration\ndescription: Configuration for BLOCK-T70.\ntype: object\nproperties:\n  sleep_for:\n    description: Duration of the sleep between opening and closing the shutter.\n    type: number\n    default: 10\n"
}
\end{lstlisting}

This block illustrates the central role that JSON-based observing blocks play in connecting high-level test definitions (managed in Jira and Zephyr Scale) with the low-level command execution carried out by the Rubin Observatory Control System. The same structure scales naturally to more complex blocks involving multiple components, on-sky targets, and longer command sequences.

\bibliography{report} 
\bibliographystyle{spiebib} 

\end{document}